\documentclass[12pt]{article}
\def\one{1\hskip-.37em 1}

\def\half{\textstyle{\frac{1}{2}}}

\def\H{{\cal H}}

\def\_{_}

\def\ra{\rightarrow}
\def\tint{{\textstyle\int}}

\def\s{\hskip.08em}
\def\d{\partial}

\def\b{\begin{eqnarray*}}  
\def\e{\end{eqnarray*}}    
\def\bn{\begin{eqnarray}}  
\def\<{\langle}
\def\>{\rangle}

\def\{{\lbrace}
\def\}{\rbrace}
\begin{document}
\title{Infinite Divisibility in \\Euclidean Quantum Mechanics}
\author{John R.~Klauder\\
Department of Physics and Department of Mathematics\\
University of Florida\\
Gainesville, FL 32611}
\date{}    
\maketitle
\begin{abstract}
In simple -- but selected -- quantum systems, the probability
distribution determined by the ground state wave function is
infinitely divisible. Like all simple quantum systems, the Euclidean
temporal extension leads to a system that involves a stochastic
variable and which can be characterized by a probability
distribution on continuous paths. The restriction of the latter
distribution to sharp time expectations recovers the infinitely
divisible behavior of the ground state probability distribution, and
the question is raised whether or not the temporally extended
probability distribution retains the property of being infinitely
divisible. A similar question extended to a quantum field theory
relates to whether or not such systems would have nontrivial
scattering behavior.
\end{abstract}
\section*{Introduction, Discussion \& Proposition}
\subsubsection*{Preliminary details}
An interesting and well studied class of quantum mechanical Hamiltonians consists of examples of the form
  \b  \H=-\s \half \d^2/\d x^2+V(x)\;,  \e
expressed in units where $m=\hbar=1$, where the real potential
$V(x)$ is chosen so that the spectrum of $\H$ is nonnegative and the
real, normalizable, nowhere vanishing ground state $\phi_o(x)$ has
zero energy eigenvalue; such systems are referred to as ``simple
systems'' in this article. In this case, the ground state itself
determines the Hamiltonian completely since
   \b  \H=-\s\half[\d^2/\d x^2-\phi''_o(x)/\phi_o(x)]\;,  \e
and therefore $\phi_o(x)$ -- just as much as $\H$ itself -- can be
said to determine all the properties of the system.

Without loss of generality we can assume that $\phi_0(x)$ is
normalized in the sense that
  \b  \int \phi_0(x)^2\,dx=1\;; \e
all integrals without indicated limits of integration extend from
$-\infty$ to $+\infty$. In that case, $\phi_0(x)^2$ determines a
probability density, and, for all real $s$, the expression
  \b  C(s)\equiv \int e^{is\s x}\s\phi_0(x)^2\,dx  \e
defines the associated characteristic function. The information
contained in the characteristic function $C(s)$ determines the
probability density $\phi_0(x)^2$ and thereby the ground state
$\phi_0(x)$ itself. Therefore, we can assert that the function
$C(s)$ determines all the properties of the system.

The distributions in an interesting {\it subclass} of all probability distributions have the property of being {\it infinitely divisible} \cite{luk}. An infinitely divisible distribution may be characterized as follows: If $X$ denotes a random variable the characteristic function of which is given by
  \b  \<e^{isX}\>\equiv C(s)=\int e^{is\s x}\s\phi_0(x)^2\,dx\;,  \e
then, for every integer $N\ge 2$, there exists $N$ independent
identically distributed random variables, $Y_n^{(N)}$, $1\le n\le N$,
such that
  \b  X=\sum_{n=1}^N\,Y_n^{(N)}\;.  \e
Alternatively stated, this property implies that the $N$th root of the characteristic function $C(s)$ is again a characteristic function, i.e.,
  \b C_{(N)}(s)\equiv C^{1/N}(s)=\int e^{isx}\,\rho_{(N)}(x)\,dx\;,  \e
where $\rho_{(N)}(x)$ denotes a probability density for all $N$. Note well that only a subclass of all probability distributions exhibits infinite divisibility.

{\bf Remark:} Integer powers of characteristic functions always
correspond again to characteristic functions.  Therefore, for
infinitely divisible distributions, the expression $C^{M/N}(s)$
again describes a characteristic functions for all positive rational numbers
$M/N$. All characteristic functions are continuous, and it follows
that in the limit as those rational numbers tend to an arbitrary positive
real number, i.e., a limit such that $M/N\ra r$, where $0<r<\infty$,
the result is again a characteristic function defined by $C^{\s
r}(s)$.

The general form of characteristic functions which are infinitely
divisible is well known \cite{luk}. For simplicity we shall confine
attention to only those distributions which are {\it even}, i.e.,
for which $V(-x)=V(x)$, leading to an even ground state
$\phi_0(-x)=\phi_0(x)$, and thereby to an even characteristic
function $C(-s)=C(s)$. In that case, the characteristic function
$C(s)$ of an even, infinitely divisible distribution may be
represented in the form \cite{luk}
  \b  C(s)=\exp\{-\half as^2-\tint[1-\cos(sy)]\,\sigma(y)\,dy\}\;,  \e
where $a\ge0$ and $\sigma(y)\ge0$ for all $y$; clearly,
$\sigma(-y)=\sigma(y)$. In this expression we have allowed ourselves
one simplification: namely, we have chosen an absolutely continuous
measure $\sigma(y)\,dy$. Existence of this expression requires that
   \b  \tint [y^2/(1+y^2)]\sigma(y)\, dy <\infty\;.  \e
However, it may well happen that
   \b  \tint \sigma(y)\,dy = \infty\;,  \e
and indeed this situation is relatively common, especially in our studies.

The general form of the characteristic function for infinitely
divisible distributions implies that the associated random variable
$X$, for which $\<e^{isX}\>=C(s)$, may be decomposed into a sum of
{\it two, independent } random variables, $X=X_G+X_P$, where the
random variable $X_G$ has a Gaussian distribution (determined by
$a$) and the random variable $X_P$ has a Poisson distribution
(determined by $\sigma$); more precisely, $X_P$ has a compound
Poisson distribution or a generalized Poisson distribution depending
on whether $\tint \sigma(y)\,dy$ is finite or infinite, respectively
\cite{fin}.

It is noteworthy that $\ln(C(s))$ is the generator of the truncated
( $=$ connected) moments of the distribution. Therefore, for
infinitely divisible distributions, it follows that
    \b  \<\s(e^{isX}-1)^T\s\>=-\half a\s s^2-\tint[1-\cos(sy)]\,\sigma(y)\,dy\;.  \e
Assuming the necessary moments exist, we learn that the truncated
moments (superscript $T$) are given by
  \b \<\s X^{2p}\s\>^T= a\s\delta_{p1}+\int y^{2p}\s\sigma(y)\,dy\;,
  \e
  and as a consequence the even order truncated moments are always
  nonnegative. This will be an important feature in our futher
  investigations.

We shall not be interested in examples that have both a Gaussian and a Poisson contribution. Indeed, we
shall focus on the wide class of examples that arise from Poisson distributions alone and thus we assume
that there is {\it no} Gaussian contribution. This means we shall set $X_G=0$ or equivalently assume that
the parameter $a$ vanishes. Having said this, we are left to focus on that subclass of characteristic
functions which can be written in the form
  \b  C(s)=\exp\{-\tint[1-\cos(sy)]\,\sigma(y)\,dy\}\;,  \e
for nonnegative functions $\sigma$ such that
  \b \tint[y^2/(1+y^2)]\,\sigma(y)\,dy<\infty\;.  \e
We shall also write
  \b U(y)^2\equiv \sigma(y)\;,   \e
and we shall call the nonnegative function $U(y)\,[=U(-y)]$ the {\it model
function}.

Just as we have asserted that the characteristic function $C(s)$
determines all the physics of our problem, we are clearly able to
declare that {\it the model function $U(y)$ determines all the
physics of our problem}. In other words, we can, without loss of
generality, adopt the model function $U(y)$ as the primary input
defining the problem at hand. Indeed, this view of the problem has
the advantage that choosing $U(y)$ initially, and within a
certain suitable class of functions, ensures that we are dealing
with an even potential $V(x)$ that leads to a ground state
$\phi_0(x)$ which in turn generates an infinitely divisible
distribution that involves only a Poisson contribution. Of course,
starting with the model function is often easier said then done!

To demonstrate that this set of conditions is not empty, it is appropriate to present a few examples.
\subsubsection*{Examples}
{\bf Example 1:} The first example is an idealized example that has the advantage of being analytically
quite simple. It is convenient to initiate our description in terms of the ground state wave function
  \b  \phi_0(x)=\frac{\sqrt{a}}{\sqrt{\pi(a^2+x^2)}}\;, \e
where $a>0$ is a fixed parameter.
In turn, this example corresponds to the potential
  \b  V(x)=\frac{2x^2-a^2}{(a^2+x^2)^2}\;.  \e
The characteristic function appropriate to this example is given by
    \b &&C(s)=\int \frac{a\s e^{isx}}{\pi(a^2+x^2)}\,dx\\
        &&\hskip.95cm=\exp(-a\s|s|)\\
        &&\hskip.95cm=\exp\{-a\s\tint[1-\cos(sy)](1/\pi\s y^2)\,dy\}\;.  \e
Thus, for this example, we see that the model function reads
  \b U(y)=\frac{\sqrt{a}}{\sqrt{\pi}}\s\frac{1}{|y|}\;.  \e

It is true that $\int U(y)^2\,dy=\infty$, due to a divergence at $y=0$. However, the present
distribution falls very slowly as $y\ra\infty$, so slowly in fact that no moments of the distribution exist.

The next example is similar to the present one except that the moments of the distribution all exist.\vskip.3cm

\noindent{\bf Example 2:}
For this model we start with the characteristic function of the ground state probability density in the form
  \b &&C(s)=e^{-b\sqrt{s^2+\rho^2}+b\rho}=\int e^{is\s x}\,\frac{b\rho}{\pi}\,
  \frac{K_1(\rho\sqrt{x^2+b^2}\s)}{\sqrt{x^2+b^2}}\s e^{b\rho}\,dx\\
&&\hskip3.8cm=\exp\{-b\int[1-\cos(s\s y)]\,\frac{\rho\s
K_1(\rho\s|y|)}{\pi\s |y|}\,dy\}\;, \e 
where $b>0$ and $\rho>0$ are free
parameters at our disposal. Here, $K_1$ denotes the standard Bessel
function. The ground state wave function for this example may be
read off from the integrand of the characteristic function. In turn,
the ground state implicitly determines the potential $V(x)$. Clearly
the model function for this case is given by
   \b  U(y)=\frac{\sqrt{b\rho\, K_1(\rho|y|)}}{\sqrt{\pi\s |y|}}\;.  \e
Since $K_1(\rho|y|)\simeq 1/(\rho|y|)$ for small argument, it follows 
that {\it near the singularity}
at $y=0$, the behavior of $U(y)$ is actually identical in the two examples 
when $b=a$; indeed, in the limit
that $\rho\ra0$, it follows that Example 2 reduces to Example 1.

As a vast generalization of both Examples 1 and 2, we may consider
characteristic functions of the general form
  \b \exp\{-b\int[1-\cos(s\s y)]\s F(y)/(\pi y^2)\,dy\}
\equiv\int e^{is\s x}\s G(x)\,dx\;,  \e
generated by basic functions $F\in C^2$ with $F(0)=1$, $F\ge0$, and
$\tint [F(y)/(1+y^2)]\,dy<\infty$. Generally speaking, after being
given an analytic expression for the basic function for such
examples, we cannot analytically specify the $L^1$ nonnegative
weight function $G$ describing the ground state probability density;
however, we know that such a probability density exists.

\subsubsection*{Euclidean time dependence}
The Hamiltonian for our system can be used to propagate the time
forward either in real time by the evolution operator $\exp(-i\H \s
T)$ or in imaginary time by the evolution operator $\exp(-\H\s T)$.
The latter case corresponds to Euclidean quantum mechanics, which as
is well known, can be described by a stochastic process involving a
stochastic variable $X(t)$, $-\infty< t< \infty$, with a
distribution determined by a Feynman-Kac like probability 
distribution \cite{bs}.
Expanding the meaning of the symbols representing expectation
values, i.e., $\<\,(\s\cdot\s)\,\>$, to cover the stochastic
variable $X(t)$, we are led to consider correlation functions of the
form \b \int\<\s X(t_1)\s X(t_2)\s\cdots\s X(t_p)\s\> \s
f_p(t_1,t_2,\dots,t_p)\,dt_1\s dt_2\s \cdots\s dt_p\;,\e for all
$p\ge1$ and all suitable weight functions $f_p$. In turn,
correlation functions may alternatively be described by a suitable
generating functional having the meaning of a characteristic
functional. Specifically, we have in mind the expression 
 \b  E\{u\}\equiv \<\, \exp[i\tint \s u(t)\s X(t)\,dt]\,\>\;,  \e
defined for all smooth test functions $u$. According to the tenants of Euclidean quantum mechanics,
the functional $E\{u\}$ obeys all the appropriate positive-definite inequalities and continuity to
satisfy the Bochner-Minlos axioms to be the functional Fourier transform of a suitable probability
distribution \cite{gelmin}.

We further observe that since the Hamiltonian is explicitly time
independent, the associated stochastic process is stationary.
Specifically this implies a time translation invariance of the
characteristic functional, which takes the form
 \b \<\,\exp[i\tint \s u(t-\tau)\s X(t)\,dt]\,\>=\<\,\exp[i\tint \s u(t)\s X(t)\,dt]\,\> \e
for any $\tau$, $-\infty<\tau<\infty$.

There is a connection of the characteristic function{al} $E\{u\}$ involving all time and the
characteristic function $C(s)$ introduced above that occurs at any one time, say at time $t=0$.
If we choose the function $u(t)$ that enters the characteristic functional as
 \b u(t)=s\,\delta(t)\;,  \e
where $\delta(t)$ is a Dirac delta function, then is follows that
  \b E\{u(\cdot)=s\,\delta(\cdot)\}=C(s)\;;  \e
hence, the restricted evaluation of the
characteristic functional $E$ coincides with the characteristic function $C$ defined at a single time.

We can make a similar connection with any one of the correlation functions themselves. For example,
let us focus on
   \b \int\s\<\s X(t_1)\cdots X(t_4)\s\>\,f_4(t_1,\ldots,t_4)\,dt_1\s\cdots\s dt_4 \e
evaluated for
  \b f_4(t_1,\ldots,t_4)=\delta(t_1)\cdots\s\delta(t_4)\;,  \e
leading to $\<\s X(0)^4\s \>$. We assert the equality given by
  \b \<\s X(0)^4\s \>=\int x^4\,\phi_0(x)^2\,dx\;,  \e
i.e., the fourth moment of the ground-state probability distribution.

\subsubsection*{Now we come to the whole point of the present paper!}
By choice, we have restricted the sharp-time, ground-state
probability distribution to be infinitely divisible. This is
encoded into the fact that an arbitrary, positive fractional power of the
characteristic function is again a characteristic function. Since
the full-time characteristic functional ($E$) collapses to the
sharp-time characteristic function ($C$) when the testing functions
are evaluated at a sharp time (say, $t=0$), it follows {\it for such
a set of limited test functions} that the sharp-time restricted
characteristic functional is infinitely divisible. The question
arises, therefore, {\it whether the FULL time characteristic
functional is itself infinitely divisible as well}, or whether that
property is restricted to those stochastic variables with a time
spread restricted to a window of finite size. It may also happen
that certain distributions enjoy full time infinite divisibility
following from sharp time infinite divisibility, while other
distributions do not exhibit full time infinite divisibility even
though they have sharp time infinite divisibility. It is clear that
the question raised here is fairly complicated. The simplest
possibility would seem to be to show that sharp time infinite
divisibility does not imply full time infinite divisibility, and
this could be determined by finding just one example for which a
single property required by infinite divisibility fails to hold.

\subsubsection*{Selection of the test question}
The structure of infinitely divisible distributions
is such that the {\it truncated} correlation functions are positive
definite functions. In particular, for such distributions, it follows that
   \b \<\s (\tint f(t)\s X(t)\,dt)^4\s\>^T\equiv\<\s
   (\tint f(t)\s X(t)\,dt)^4\s\>-3\s\<\s (
   \tint f(t)\s X(t)\,dt)^2\s\>^2>0 \e
unless $f=0$. The simplest example derives from a uniform weighting
where $f(t)=1$ in an interval $-T<t<T$, and $f(t)$ is zero
elsewhere. We should like to examine the behavior of this truncated
correlation function for asymptotically large $T$, and compare it
with the behavior for extremely small $T$. Due to stationary of the
underlying ensemble, the large $T$ behavior diverges as $2T$; for
small times, on the other hand, it behaves as $(2T)^4$ times the
sharp time expectation value. To eliminate these unimportant
temporal factors, it is convenient to rescale the truncated four
point function by a factor $(1+2T)^3/(2T)^4$, which leads to the
alternative form given by
  \b &&\chi_2\equiv\frac{(1+2T)^3}{(2T)^4}\,\bigg[\<\s (\int_{-T}^T
  X(t)\,dt)^4\s\>^T\s\bigg]\\
  &&\hskip.59cm=\frac{(1+2T)^3}{(2T)^4}\,\bigg[\<\s (\int_{-T}^T
  X(t)\,dt)^4\s\> -3\<\s (\int_{-T}^T X(t)\,dt)^2\s\>^2\bigg]\;. \e

Although this is just one measure of the truncated four point
function it is a significant one and one that is comparatively easy
to evaluate. If $\chi_2<0$ for large $T$, it would establish that
the full time probability functional is {\it not} infinitely
divisible even though the sharp time probability function has been
chosen (by design) to be infinitely divisible. Of course, if
$\chi_2>0$ for all $T$, then no definitive conclusion can be drawn
about the nature of the full time probability distribution. To
proceed further, one would need to check other moments, or even
change to another model function in order to test the nature of the
full time probability distribution. At present, the author is not
aware of any general scheme that could decide whether or not the
full time probability functional is infinitely divisible whenever
the sharp time probability function has been chosen to be infinitely
divisible. Nevertheless, in the absence of any proof to the
contrary, it seems reasonable to {\it conjecture} that infinite
divisibility for sharp time simple quantum systems does {\it not} imply
full time infinite divisibility for them.

\subsubsection*{Alternative representation}
It is noteworthy that the evaluation of $\chi_2$ can be reexpressed
in terms of the eigenfunctions and eigenvalues of the Hamiltonian
for our simple system, and we now turn our attention to develping
this alternative expression for $\chi_2$. For convenience, let us
consider those special cases where $\H$ has a purely discrete,
nondegenerate spectrum such that
  \b [\s-\s \half \d^2/\d x^2+V(x)\s]\phi_n(x)=E_n\s \phi_n(x)\;,  \e
where $0=E_0<E_1<E_2\cdots$ and all the (real) eigenfunctions form a
complete orthonormal set of functions for which
  \b \tint \phi_n(x)\s\phi_m(x)\,dx=\delta_{nm}\;.  \e
 Here, as before, the ground state is denoted by $\phi_0(x)$. We shall also introduce
 these relations in the usual abstract bra-ket notation as well, namely
  \b  [\s \half \s P^2+V(Q)\s]\s|n\>=E_n\s|n\>\;,  \e
with $|n\>$ denoting the abstract eigenvectors, for which
$\<n|m\>=\delta_{nm}$. Of course, we let $|0\>$ denote the ground
state.

 By definition, the correlation functions are {\it symmetric}
functions of their temporal arguments. However, when expressed in terms of
the eigenvectors and eigenvalues, it proves convenient to define the
correlation functions in a {\it time ordered} manner. In particular,
for the time ordered two point function, this rule leads to the
expression
   \b  \<\s X(t)\s X(u)\s\>=\sum_{n=0}^\infty \<0|Q|n\>\s e^{-E_n(t-u)}
   \s\<n|Q|0\>\;,\hskip1cm t\ge u\;; \e
we can of course extend this expression to all $t$ and $u$ values
simply by replacing $(t-u)$ by $|t-u|$ on the right hand side, but
we choose not to do so in order to simplify the analysis in what
follows.
 It is clearly convenient to introduce the notational
shorthand that
     \b   Q_{kl}\equiv\<k|\s Q\s|l\>\;.  \e
Similar expressions also exist for higher-order correlation
functions. For example, for the time ordered four point function we
have
  \b &&\<\s X(t)\s X(u)\s X(v)\s X(w)\s\>=\sum_{k,l,m=0}^\infty\,Q_{0k}
  \s e^{-E_k(t-u)}\s Q_{kl}\s e^{-E_l(u-v)}\\
&&\hskip4cm\times\s Q_{lm}\s e^{-E_m(v-w)}\s Q_{m0}\;, \hskip1cm
t\ge u\ge v\ge w\;. \e

From the assumed symmetry of the potential, i.e., $V(-x)=V(x)$, it
follows that the eigenfunctions have {\it alternating parity},
namely, that $\phi_n(-x)=(-1)^n\s\phi_n(x)$. Consequently, the
matrix elements of $Q$, such as $\<k|Q|l\>=Q_{kl}$, only connect
eigenvectors of {\it opposite} parity; in other words, if $k$ is
even, then $l$ is odd, or vice versa. In particular, for the two
point function, as presented above, only {\it odd} values of $n$
lead to nonvanishing contributions. For the four point function, as
presented above, only odd values of $k$ and $m$ contribute, while
only even values of $l$ need be considered.

To begin our construction of $\chi_2$, we may integrate our time
ordered expressions over a restricted time region and rescale the
result to account for such a limited integration domain. Initially,
this recipe implies that
  \b &&\<(\int_{-T}^T X(t)\,dt)^4\s\>=24\,\bigg[\sum_{k,l,m} Q_{0k}\,Q_{kl}
  \,Q_{lm}\,Q_{m0}\bigg]\,
  \int_{-T}^Tdt\int_{-T}^tdu \\
 &&\hskip1cm\times\int_{-T}^udv \int_{-T}^vdw
 \exp[-E_k(t-u)-E_l(u-v)-E_m(v-w)]\;.  \e
Here the factor $24=4!$ corrects for the limited domain of
integration due to time ordering. For small $T$, the integral above
involves terms $O(T^4)$, while for large $T$, this integral involves
terms $O(T^2)$, $O(T)$, and $o(T)$. Since we content ourselves with
the integral's value for very small and very large $T$ values, it
will suffice to consider terms of the indicated types, respectively.

\subsubsection*{Small $T$ behavior}
 First, consider very small $T$ values, or more
specifically consider the situation where $E_k\s T\ll1$ for all
sensibly contributing energy values. In that case, we can replace
all exponentials in the previous formula by unity, and therefore, to
leading order
   \b  \frac{(1+2T)^3}{(2T)^4}\,\<(\int_{-T}^T X(t)\,dt)^4\s\>=\sum_{k,l,m} Q_{0k}\,Q_{kl}
  \,Q_{lm}\,Q_{m0}\;,\hskip.5cm T\ll1\;.  \e
This result of course is equivalent to the sharp time four point
moment. Second, consider the square of the two point function, which
leads to the result
   \b  \frac{(1+2T)^3}{(2T)^4}\,\<(\int_{-T}^T
 X(t)\,dt)^2\>^2=\sum_{k,m=1}^\infty\,Q_{0k}\s Q_{k0}\s Q_{0m}\s Q_{m0}\;,\hskip.5cm T\ll1\;,\e
 to leading order in $T$. This expression, of course, is just the
 square of the sharp time second moment. Finally, the truncated
 four point moment for small $T$ is given by
 \b  \chi_2=\sum_{k,l,m=1}^\infty\,Q_{0k}\s Q_{kl}\s Q_{lm}\s
 Q_{m0}-2\sum_{k,m=1}^\infty\,Q_{0k}\s Q_{k0}\s Q_{0m}\s Q_{m0}\;;
 \e
 note well, that for the first term all three sums omit the ground
 state, and for the second term both sums omit the ground state.
 This change of the first term has resulted in the factor $3$
 becoming a factor $2$ in the second term.

 We can readily check this
 result for the harmonic oscillator, namely where $V(x)=\half\omega^2\s x^2$.
 In that case, the only nonvanishing matrix elements of interest
 are $Q_{01}=Q_{10}=1/\sqrt{2\s\omega}$ and $Q_{12}=Q_{21}=1/\sqrt{\omega}$. Therefore,
   \b &&\chi_2=Q_{01}\s Q_{12}\s Q_{21}\s Q_{10}-2\s [\s Q_{01}\s
   Q_{10}\s]^2\\
   &&\hskip.55cm=[1/(2\s\omega^2) -2/(4\s\omega^2)]=0\;.  \e
This is the correct result: Since the ground state for the harmonic
oscillator is a Gaussian, the distribution is infinitely divisible;
and, as a Gaussian, all truncated moments other than the second
moment vanish.

\subsubsection*{Large $T$ behavior}
Let us now turn our attention to the evaluation of $\chi_2$ for
large $T$. In this case it is useful to divide the fourth moment
into two distinct expressions, namely,
  \b \frac{(1+2T)^3}{(2T)^4}\,\<\s(\int_{-T}^T X(t)\,dt)^4\s\>=
  A+B\;. \e
  In this expression,
\b A\equiv 24\,\sum_{k,l,m=1}^\infty\,Q_{0k}\s Q_{kl}\s Q_{lm}\s
  Q_{m0}\,\bigg[\s\frac{1}{E_k}\,\frac{1}{E_l}\,\frac{1}{E_m}\s\bigg]\;,  \e
  which is valid for $T\gg1$, provided all $E$ values are strictly
  positive, $E>0$. This restriction leads to the omission of the term for
  $l=0$ above. The omitted term, where $l=0$, can be evaluated
  separately, and to leading order it follows that
    \b B\equiv
24\,\sum_{k,m=1}^\infty\,Q_{0k}\s Q_{k0}\,Q_{0m}\s Q_{m0}\s\bigg[\s
T\s\frac{1}{E_m}\,\frac{1}{E_k}-\frac{1}{E_k^2}\, \frac{1}{E_m}-
\frac{1}{E_k}\, \frac{1}{E_m^2}\s\bigg]\;.\e
The two point function
for large $T$ follows from the expression given by
   \b \<\s(\int_{-T}^T X(t)\,dt)^2\s\>=\sum_{m=1}^\infty\,Q_{0k}\s Q_{k0}\s
   \bigg[\,\frac{4T}{E_k}-\frac{2}{E_k^2}\,\bigg]\;.\e
Consequently, for very large $T$, we observe that
   \b &&\frac{(1+2T)^3}{(2T)^4}\bigg(\<\s(\int_{-T}^T
   X(t)\,dt)^2\s\>\bigg)^2\\
   &&\hskip1cm=\frac{1}{2T}\sum_{k,m=1}^\infty\,Q_{0k}\s
   Q_{k0}\,Q_{0m}\s Q_{m0}\,
   \bigg[\,\frac{4T}{E_k}-\frac{2}{E_k^2}\,\bigg]\bigg[\,\frac{4T}{E_m}-\frac{2}{E_m^2}\,\bigg]\e
Finally, for large $T$,
  \b &&\frac{(1+2T)^3}{(2T)^4}\,\<(\int_{-T}^T X(t)\,dt)^4\>^T\\
    &&\hskip1cm=24\sum_{k,l,m=1}^\infty\,Q_{0k}\,Q_{kl}\,Q_{lm}\,Q_{m0}\,\bigg[\frac{1}{E_k}\,\frac{1}{E_l}\,\frac{1}{E_m}\bigg]\\
&&\hskip1cm+24\,T\sum_{k,m=1}^\infty\,Q_{0k}\,Q_{k0}\,Q_{0m}\,Q_{m0}\,\bigg[\frac{1}{E_k}\,\frac{1}{E_m}\bigg]\\
  &&\hskip1cm-24\,\sum_{k,m=1}^\infty\,Q_{0k}\s Q_{k0}\,Q_{0m}\s Q_{m0}\,
  \bigg[\s\frac{1}{E_k}\s\frac{1}{E_m^2}+\frac{1}{E_k^2}\s\frac{1}{E_m}\s\bigg]\\
  &&\hskip1cm-3\,\frac{1}{2T}\sum_{k,m=1}^\infty\,Q_{0k}\s
   Q_{k0}\,Q_{0m}\s Q_{m0}\,
   \bigg[\,\frac{4T}{E_k}-\frac{2}{E_k^2}\,\bigg]\bigg[\,\frac{4T}{E_m}-\frac{2}{E_m^2}\,\bigg]
   \;,\e
   which to leading order in $T$ becomes
\b &&\chi_2=24\,\sum_{k,l,m=1}^\infty\,Q_{0k}\,Q_{kl}\,Q_{lm}\,
Q_{l0}\,\bigg[\frac{1}{E_k}\,\frac{1}{E_l}\,\frac{1}{E_m}\,\bigg]\\
  &&\hskip2cm-24\,\sum_{k,m=1}^\infty\,Q_{0k}\s Q_{k0}\,Q_{0m}\s Q_{m0}
  \,\bigg[\s\frac{1}{E_k}\s\frac{1}{E_m^2}\s\bigg]\;.\e

Again, it is useful to test this expression with the harmonic
oscillator. Besides the matrix elements already given previously, we
need the first two excited state energy levels, $E_1=\omega$ and
$E_2=2\s\omega$. In that case, we find that
  \b
  \chi_2=24\,[(1/2\s\omega^2)\cdot(1/2\s\omega^3)-
(1/4\s\omega^2)\cdot(1/\omega^3)]=0\;,
  \e
  as expected.

\subsubsection*{Summary of principal formulas}
The truncated four point function given by
  \b  \chi_2=(1+2T)^3/(2T)^4\,\<(\int_{-T}^T X(t)\,dt)^4\>^T  \e
  is expressed alternatively by
    \b \chi_2=\sum_{k,l,m=1}^\infty\,Q_{0k}\s Q_{kl}\s Q_{lm}\s
 Q_{m0}-2\sum_{k,m=1}^\infty\,Q_{0k}\s Q_{k0}\s Q_{0m}\s Q_{m0}\;,
 \e
 for {\it small} $T$, and by
   \b  &&\chi_2=24\,\sum_{k,l,m=1}^\infty\,Q_{0k}\,Q_{kl}\,Q_{lm}\,
Q_{l0}\,\bigg[\frac{1}{E_k}\,\frac{1}{E_l}\,\frac{1}{E_m}\,\bigg]\\
 &&\hskip2cm
-24\,\sum_{k,m=1}^\infty\,Q_{0k}\s Q_{k0}\,Q_{0m}\s Q_{m0}
  \,\bigg[\s\frac{1}{E_k}\s\frac{1}{E_m^2}\s\bigg]\;,\e
for {\it large} $T$. In this expression, $Q_{kl}\equiv\<k|Q|l\>$
denotes a matrix element of the position operator $Q$, where the
states $|k\>$ are normalized eigenvectors of the time-independent
Schr\"odinger equation,
   \b [\s\half\s P^2+V(Q)\s]\,|k\>=E_k\s|k\>\;,  \e
for the real, even potential, $V(Q)$.

{\bf Remark:} The apparent difference in dimensions between the
behavior for small and large $T$ arises from our choice of a scaling
factor; had we used
  \b  (1+2E T)^3/(2ET)^4  \e
  instead of $(1+2T)^3/(2T)^4$, where $E$ is some characteristic
  energy level, then the two extremal expressions would have had the
  same dimensions. However, since our main concern is to study
  simple systems for which $\chi_2>0$ for small $T$ with the aim of
  finding examples for which $\chi_2<0$ for large $T$, the formal
  difference in dimensionality is of little concern.

\subsubsection*{Why bother?}
The reader may well ask why should one care about quantum systems
that have infinitely divisible distributions for sharp time position
variables and may -- or may not -- have infinitely divisible
distributions for full time position variables. A brief
explanation may help clarify the situation.

A Euclidean formulation for a scalar quantum field theory is
characterized by a stochastic field variable we may call
$\phi(x,t)$. The statistics of this variable are governed by a
probability distribution, which, like our single degree of freedom
examples discussed above, may be described by a characteristic
functional
   \b \<\s e^{i\tint u(x,t)\s\phi(x,t)\,dx\s dt}\s\>\;. \e
The sharp time expression is given simply by setting
$u(x,t)=v(x)\s\delta(t)$ leading to
  \b   \<\s e^{i\tint v(x)\s\phi(x,0)\,dx}\s\> \;. \e

Interacting quantum field theories encounter divergences, and this
is as true for the sharp time field expressions as it is for the
full time field expressions. In a recent paper \cite{kla6} it was
argued that sharp time formulations involving an infinitely
divisible field distribution tended to alleviate some of the
principal causes of field theory divergences. From this point of
view there would seem to be some merit in seeking to formulate
matters so that sharp time fields had infinitely divisible
distributions. Accepting such an argument opens the question of
whether an infinitely divisible distribution for sharp time fields
does -- or does not -- force the full time field distribution to be
infinitely divisible.

Suppose, for the sake of argument, it was true that the full
time field distribution was also infinitely divisible. By an
argument of Buchholz and Yngvason \cite{buch} this would result in a
theory with a unit scattering matrix. For example, consider the
truncated four point function
   \b  \<\s\phi(f_1)\s\phi(f_2)\s\phi(g_2)\s\phi(g_1)\s\>^T\;,  \e
   where
   \b \phi(f)\equiv \tint f(x,t)\s\phi(x,t)\,dx \s dt  \;,  \e
for $f$ a smooth test function, etc. For infinitely divisible 
distributions certain truncated
correlation functions are nonnegative, such as
  \b  0\le\<\s\phi(g_1)\s\phi(g_2)\s\phi(g_2)\s\phi(g_1)\s\>^T \;.
  \e
Consequently, by the Schwarz inequality,
  \b &&\hskip-.8cm0\le|\<\s\phi(f_1)\s\phi(f_2)\s\phi(g_2)\s\phi(g_1)\s\>^T|^2\\
  &&\le\<\s\phi(f_1)\s\phi(f_2)\s\phi(f_2)\s\phi(f_1)\s\>^T
  \<\s\phi(g_1)\s\phi(g_2)\s\phi(g_2)\s\phi(g_1)\s\>^T\;.  \e
Now, take suitable limits so that
  \b
  &&\phi(g_1)\ra\phi_{in}(g_{1})\;\;,\hskip1cm\phi(g_2)\ra\phi_{in}(g_{2})\;\;,\\
  &&\phi(f_1)\ra\phi_{out}(f_{1})\;,\hskip1cm\phi(f_2)\ra\phi_{out}(f_{2})\;,
  \e
where ``$in$'' and ``$out$'' fields denote free fields appropriate to 
the asymptotic regime. In that case we find
\b &&\hskip-.3cm0\le|\<\s\phi_{out}(f_{1})\s\phi_{out}(f_{2})\s
\phi_{in}(g_{2})\s\phi_{in}(g_{1})\s\>^T|^2\\
  &&\le\<\s\phi_{out}(f_{1})\s\phi_{out}(f_{2})\s\phi_{out}(f_{2})\s\phi_{out}(f_{1})\s\>^T \\
  &&\hskip.5cm\times\<\s\phi_{in}(g_{1})\s\phi_{in}(g_{2})\phi_{in}(g_{2})\s\phi_{in}(g_{1})\s\>^T=0\;,  \e
leading to no two particle scattering. Extending this type of
argument leads to a trivial theory in the sense that the scattering
matrix $S=\one$.

If, however, an infinitely divisible sharp time field distribution
did {\it not} imply that the full time field distribution was
infinitely divisible, then triviality of the scattering matrix may
be avoided.

This kind of question is extremely hard to study for a field theory.
Hence, the study initiated in this paper is restricted to a single
degree of freedom quantum system with the Hamiltonian form chosen
to be similar in spirit to that of a traditional scalar field
theory. If a positive outcome of the single degree of freedom
problem emerges, it may well suggest that further study of the
quantum field situation may be worthwhile.

\subsubsection*{A proposed  problem}
As a simple examination of the associated ground state reveals, 
neither of the two specific examples (Examples 1 \& 2) 
suggested above 
are suitable for a detailed study since their spectrum is part discrete 
and part continuous. Unfortunately, distribution functions for general 
infinitely divisible characteristic functions are known for very 
few examples \cite{luk}. We conjecture that certain examples, such as 
\b && C(s)=\exp\{-\tint[1-\cos(sy)]\,e^{-y^2}/(\pi\s |y|^\alpha)\,dy\}\\
   &&\hskip.96cm \equiv\int \cos(sx)\,\phi_0(x)^2\,dx\;,  \e
for $2\le\alpha<3$, may correspond to potentials with a purely 
discrete spectrum which would then permit our formulas to be applied. 

The implicitly defined ground state $\phi_0(x)$ above may be determined 
numerically. If the ground state falls to zero suitably faster than 
an exponential, then the spectrum 
of the Hamiltonian would be purely discrete. In turn, the potential 
associated with this ground state is
   \b  V(x)\equiv \phi''_0(x)/2\s\phi_0(x)\;, \e
which again could be numerically determined.  Even if there are small 
errors in the numerical determination, it is quite likely that the simple 
system still lies within the special class that is infinitely divisible for
the ground state distribution. With the potential $V(x)$ -- with 
symmetry enforced -- now determined, standard computer programs could 
be used to 
calculate a number of eigenfunctions and eigenvalues, as well 
as suitable matrix elements. These quantities permit the calculation 
of $\chi_2$ for large and small $T$ values. If, for large $T$, $\chi_2<0$,
the desired result will have been achieved; if, instead, $\chi_2>0$, then
further investigations are warranted. 

Of course, it may be true that sharp time infinite divisibility necessarily
implies full time infinite divisibility. If this implication could be 
proved, then that would be an important result which would make any 
computation (such as outlined in the previous paragraph) 
completely unnecessary.

\end{document}